\documentclass[doublecol]{epl2}

\usepackage{graphicx}
\usepackage{dcolumn}
\usepackage{bm}
\usepackage{stmaryrd}
\usepackage{latexsym}
\usepackage{amssymb}
\usepackage{amsfonts}
\usepackage{amsmath}
\usepackage{color}

\title{Pairing Symmetry in Iron-Pnictide Superconductor KFe$_2$As$_2$}

\author{Wei Li\inst{1,2}, Jian Li\inst{2}, Jian-Xin Zhu\inst{3}, Yan Chen\inst{1}, and C. S. Ting\inst{2}}
\institute{
\inst{1}Department of Physics and State Key Laboratory of Surface Physics, Fudan University, Shanghai 200433, China\\
\inst{2}Texas Center for Superconductivity and Department of Physics, University of Houston, Houston, Texas 77204, USA\\
\inst{3}Theoretical Division, Los Alamos National Laboratory, Los Alamos, New Mexico 87545, USA
.}
\abstract{The pairing symmetry is one of the major issues in the study of iron-based superconductors. We adopt a ten-orbital model by using the maximally localized Wannier functions based on the first-principles band structure calculations combining with the $J_1$-$J_2$ model for KFe$_2$As$_2$, the phase diagram of pairing symmetries is constructed. We find that the pairing symmetry for KFe$_2$As$_2$ is a nodal $s_{x^2y^2}+s_{x^2+y^2}$-wave in the folded Brillouin zone with two iron atoms per unit cell. This pairing symmetry can explain the experiments observed nodes, and it also can be tested by future experiments.}


\begin{document}


\maketitle

The discovery of the superconducting iron-pnictide material (1111-type $Re$OFeAs, $Re$=rare earth)\cite{ref1} in 2008 has triggered great research interests which led to synthesize similar iron-based superconductors, such as 122-type $B$Fe$_2$As$_2$ ($B$=Ba, Sr, or Ca)\cite{ref2}, 111-type $A$FeAs ($A$=alkali metal)\cite{ref3}, 11-tpye $\alpha$-FeSe(Te)\cite{ref4}, and new-type K$_x$Fe$_{2-y}$Se$_2$\cite{ref5} materials. Their crystal structures are tetrahedral with the divalent iron square planes. A characteristic feature of these superconducting systems is that the band structure near the Fermi energy is derived from Fe-$3d$ orbitals with only modest hybridization of ligand-$p$ orbitals\cite{ref6}. Most of these compounds were reported to show superconductivity after doping or under high pressure and have the same robust tetrahedral layer structure.

To uncover the mechanism of superconductivity in these materials, the determination of pairing symmetry of the superconducting order parameter is a good starting point. In most of the electron-doped and weakly hole-doped 1111 and 122 compounds, the band structure calculations show that there are disconnected quasi-two dimensional (2D) hole and electron pockets\cite{ref7,ref8}. The former ones are centered at the $\Gamma$ point, while the latter ones are located at the $M$ point of the Brillouin zone (BZ) of two iron atoms per unit cell. Strong scattering between the electron and hole pockets corresponding to a nesting wavevector\cite{Cvetkovic2009}: $q\sim(\pi,\pi)$, which could lead to the superconductivity with the so-called $S^{\pm}$ pairing symmetry\cite{ref7}. The order parameter with this pairing symmetry is nodeless, and has a sign changed from the hole to the electron pocket. Such a $S^{\pm}$ pairing symmetry is gaining increasing experimental supports\cite{HDing2008,Hanaguri}. For a heavily hole-doped compound like KFe$_2$As$_2$, the angle-resolved photoemission spectroscopy (ARPES) measurements\cite{HDing,Yoshida} indicated that the electron pockets near $M$ points were replaced by the ellipse-like hole pockets while the hole pockets near the $\Gamma$ point became larger and closed to $\pi/2$ in the folded BZ of two iron atoms per unit cell. The superconducting transition temperature of this compound is $T_c\sim 3.6$ K which is much lower than those in other iron-based superconductors. Under this situation $q\sim(\pi,\pi)$ is no longer the nesting wavevector and the superconductivity pairing interactions could come from small $q$ scattering. Although both experimental observations\cite{Furukawa,JKDong,Terashima,Reid,Hashimoto,Fukazawa,SWZhang,Ohishi} and theoretical investigations\cite{Thomale,Suzuki,Maiti,SMaitiPRL2011,SMaitiPRB2011} give nodal gap structures in KFe$_2$As$_2$, the pairing symmetry of this compound is still controversial. Specifically, experiments on small-angle neutron scattering\cite{Furukawa} indicates nodal lines perpendicular to the $c$-axis while experiments on thermal conductivity\cite{JKDong,Reid} and superfluid density\cite{Hashimoto} show a $d$-wave like nodal structure. Theoretically, based on the weak
coupling approaches, a functional renormalization group study shown the dominant $d$-wave state\cite{Thomale} is most favorable, while Fermi surface (FS) restricted random phase approximation type spin-fluctuations analysis\cite{Suzuki} and an analytical study in which the interactions are approximated by their lowest angular harmonics\cite{SMaitiPRL2011,SMaitiPRB2011} have revealed that $s$-wave and $d$-wave pairing amplitudes have near-equal strength. An $s$-wave state with accidental line nodes that run either parallel to the $c$ axis is also proposed by S. Maiti {\it et al.}\cite{Maiti}.

In this paper, we address the pairing symmetry in KFe$_2$As$_2$ from a strong coupling approach, and treat the competing pairing channels on the same footing. Based on this approach, it has been shown that the pairing symmetry is determined mainly by the magnetic exchange coupling, as well as the FS topology for other iron-based materials\cite{KSeo,Goswami,RYu,YHSu}. In this sense, we construct a ten-orbital model by using the maximally localized Wannier functions based on the first-principles band structure calculations and combine with the $J_1$-$J_2$ model to discuss all possible pairing symmetry. The mean-field phase diagram of a $t$-$J_1$-$J_2$ model with correct band structure for this compound is obtained by performing the self-consistently calculations. We find that only the $s_{x^2+y^2}+s_{x^2y^2}$-wave can give rise to the nodal like structure in the folded BZ with two iron atoms per unit cell, and only this wave pairing symmetry can explain the experiments which evidenced a nodal order parameter\cite{JKDong,Reid,Hashimoto}. This result can be tested by future experiments, such as APRES measurements.

\begin{figure}[tbp]
\includegraphics[bb=15 1 730 410, width=8cm, height=4.3cm]{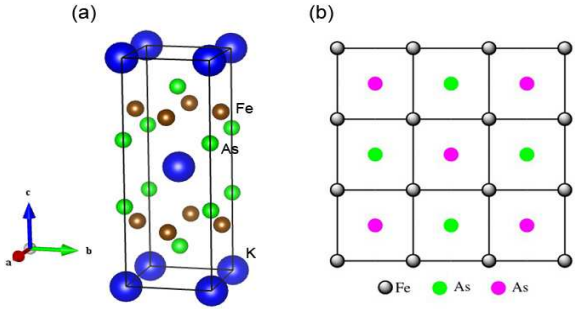}
\caption{(Color online) (a) The schematic crystal structure of KFe$_2$As$_2$; (b) Sketch of the lattice structure of FeAs layers. The As ions {\textcolor{green}\textbullet} and {\textcolor{magenta}\textbullet} are
located just above and below the center of each face of the Fe square lattice, respectively.}\label{fig:fig0}
\end{figure}

In the first-principles calculations, the plane wave basis method is implemented in the vasp code\cite{VASP}, and the Perdew-Burke-Ernzerhof exchange correlation potential\cite{PBE} has been used. A 500 eV cutoff in the plane wave expansion and a $12\times 12\times 10$ Monkhorst-Pack k-grid are chosen to ensure the calculation with an accuracy of $10^{-5}$ eV, and all structures (lattice constants as well as internal coordinates) were optimized until forces on individual atoms were smaller than 0.005 eV/\AA\ to obtain sufficient accuracy throughout the calculations. The crystal structure is shown in Fig. \ref{fig:fig0}(a).

\begin{figure}[tbp]
\includegraphics[bb=5 1 790 390, width=8.5cm, height=4.3cm]{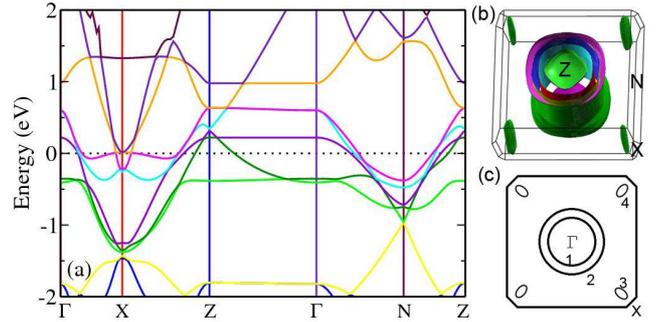}
\caption{(Color online) (a) Energy band structure in the NM state of KFe$_2$As$_2$, and (b) the corresponding FS sheets crossing the Fermi energy in the band structure; (c) 2D cross-sectional ($k_z$=0) representation of FS of the band structure calculations. The Fermi energy is set to zero.}\label{fig:fig1}
\end{figure}

Firstly, we focus on the electronic band structure, as shown in Fig. \ref{fig:fig1}. It includes the calculated band structure (a), the corresponding FS (b) and the 2D cross-section of the FS at $k_z$=0 (c) in the nonmagnetic (NM) state in the body-centered tetragonal unit cell. These results are similar to the experimental observations of this material\cite{HDing,Hashimoto,TERASHIMA,Yoshida,Suzuki}. As we can see in Fig. \ref{fig:fig1}, there are six FS pockets, in which the three hole pockets (two of them are degenerate) are very cylindrical, suggesting a strong 2D behavior. However, there is another pocket near $Z$ point with highly three-dimensional characteristics but vanishing around $\Gamma$ point. In addition, four small ellipse-like hole pockets located at around the $X$ points of the BZ. This result is consistent with ARPES results and previous first-principles calculations\cite{HDing,Yoshida}. Although these results have been reported, we present them here to facilitate the discussion on the orbitals characters and the possible pairing symmetry in superconducting state based on this FS topology.

\begin{figure}
\begin{center}
\tabcolsep=-0.5cm
\begin{tabular}{c}
\includegraphics[bb=50 50 720 560,width=4.0cm,height=3.2cm]{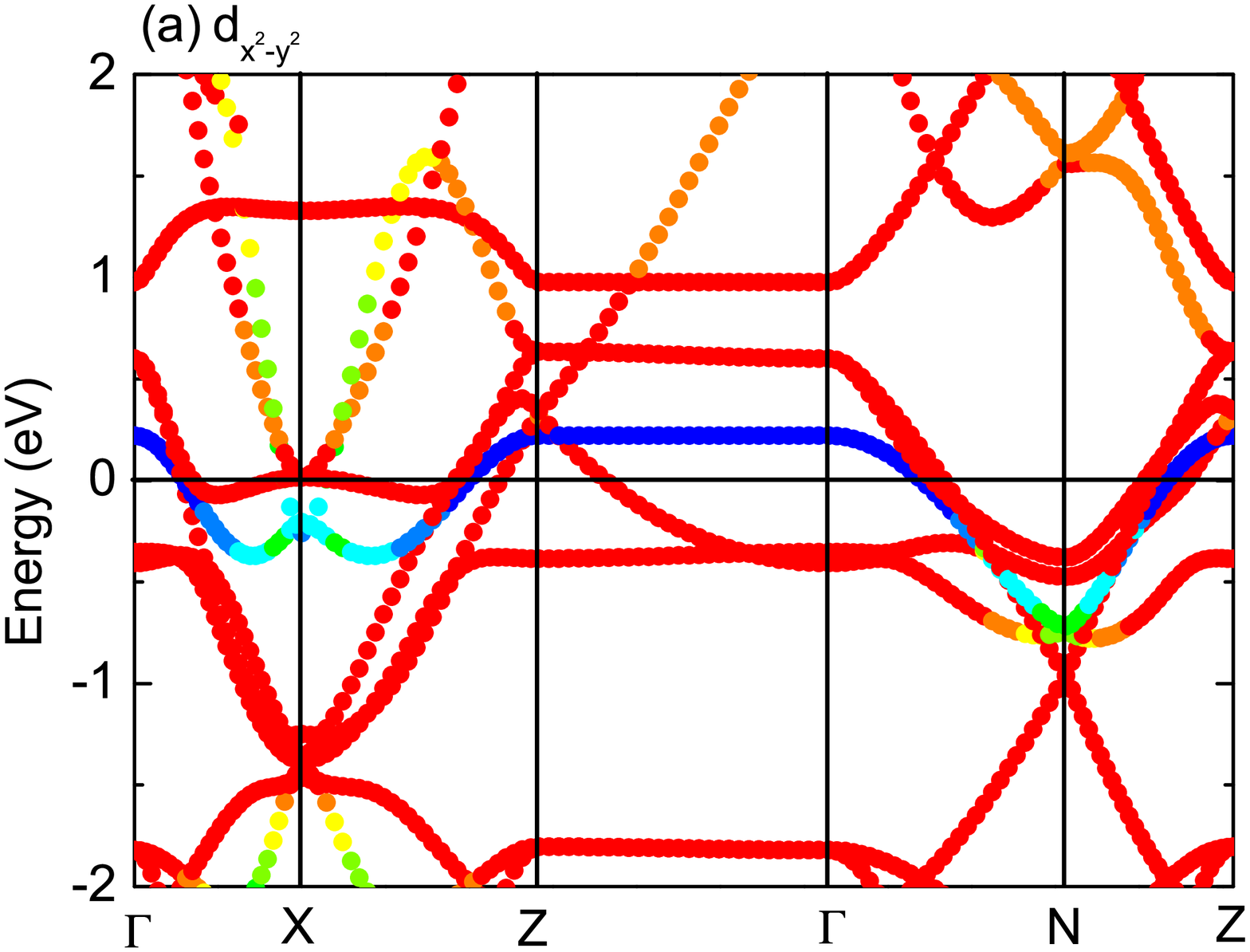}
\includegraphics[bb=50 50 820 560,width=4.5cm,height=3.2cm]{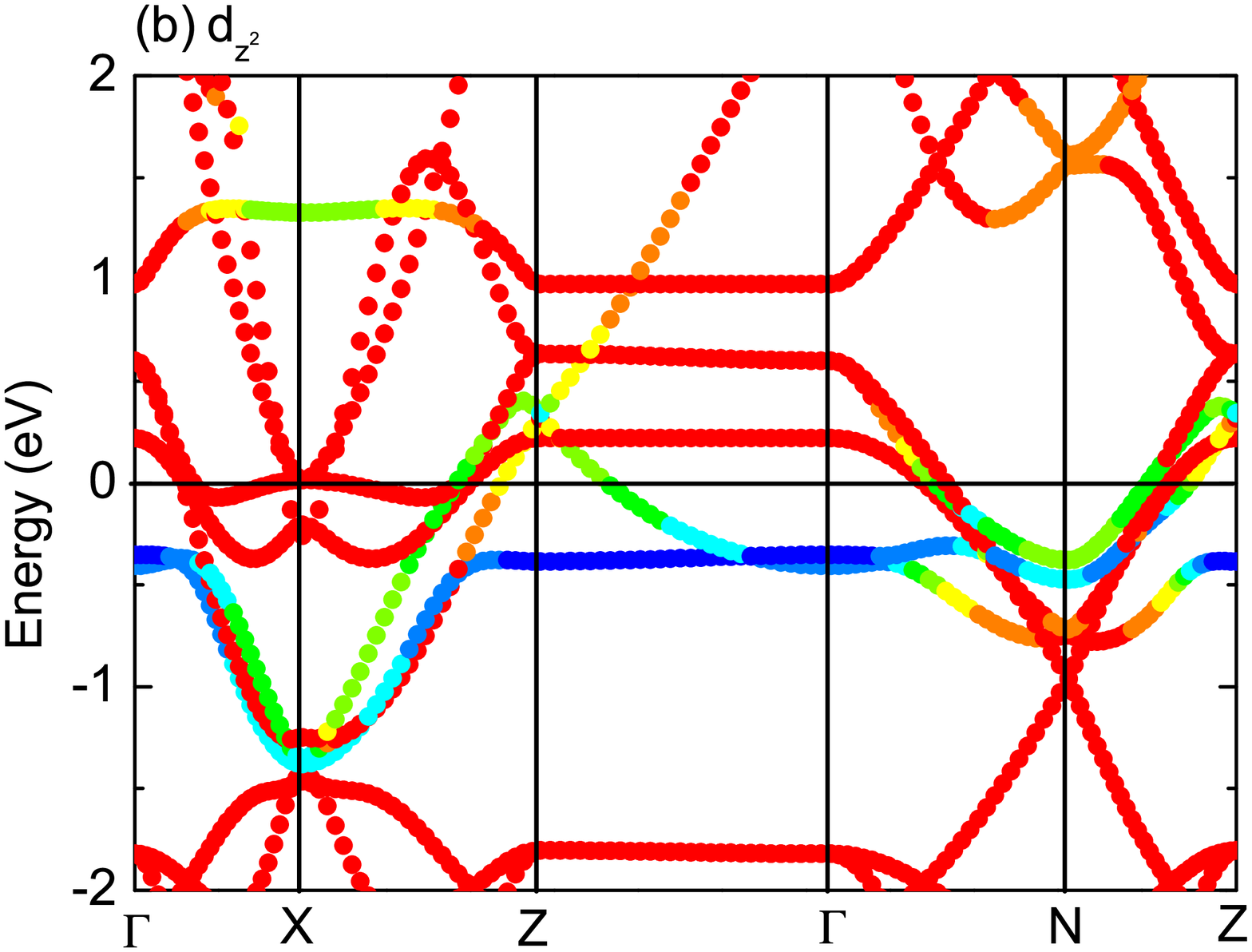}\\
\includegraphics[bb=50 50 720 560,width=4.0cm,height=3.2cm]{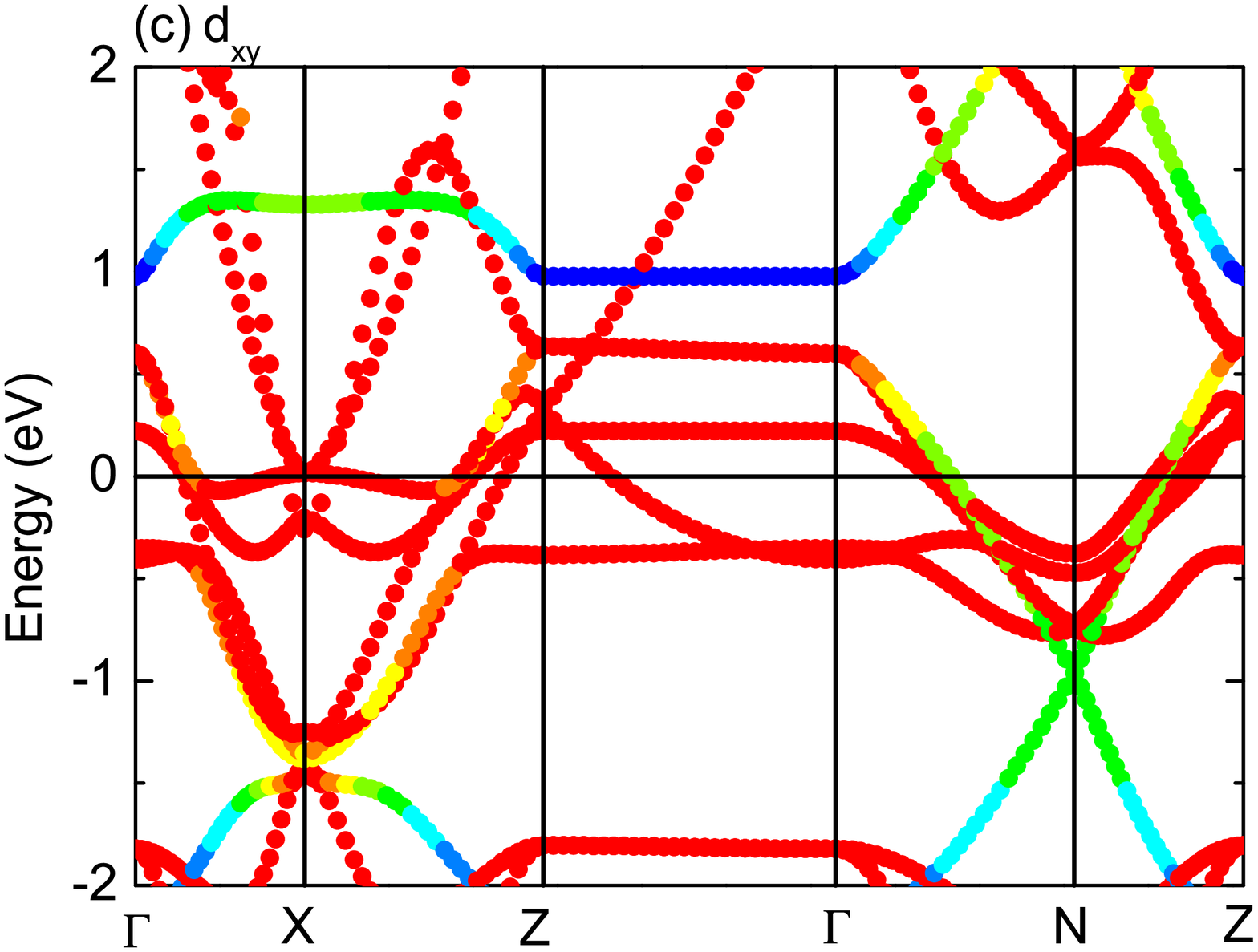}
\includegraphics[bb=50 50 820 560,width=4.5cm,height=3.2cm]{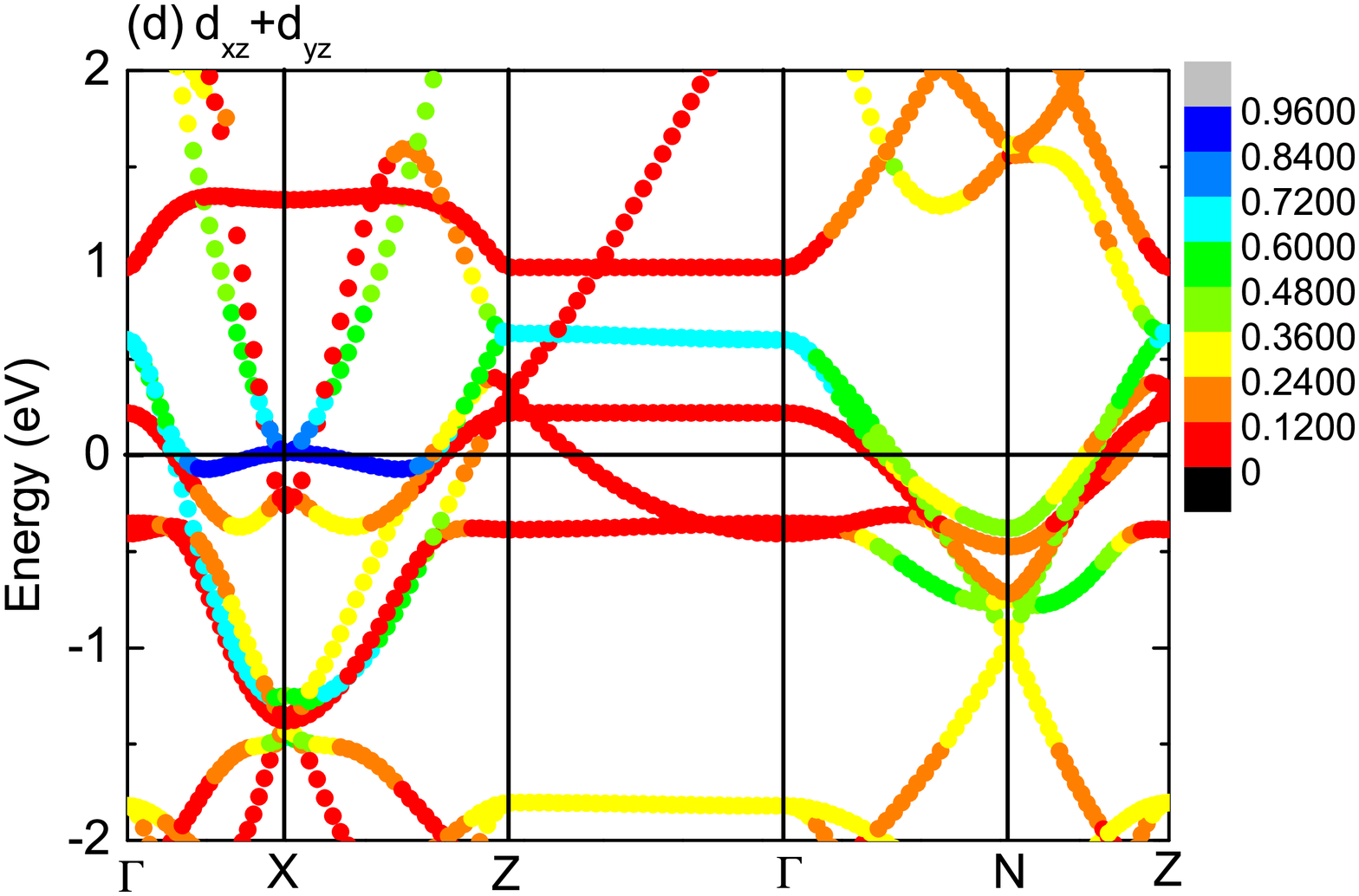}
\end{tabular}
\end{center}
\caption{(Color online) Band structure of KFe$_2$As$_2$, decorated with partial characters of the $e_g$ [top, $d_{x^2-y^2}$ (a) and $d_{z^2}$ (b)] and $t_{2g}$ [bottom, $d_{xy}$ (c) and $d_{xz}+d_{yz}$ (d)] of the Fe-$3d$ bands. The Fermi energies are all set to zero.}\label{fig:fig2}
\end{figure}

In order to clarify the orbitals characters in each pocket, we project the bands onto the five Fe-$3d$ orbitals shown in Fig. \ref{fig:fig2}. From Fig. \ref{fig:fig2}, we can see that there is a double degenerate cylindrical hole pocket at $\Gamma$ point with a dominant Fe-$3d_{xz}$ and Fe-$3d_{yz}$ orbitals characters. However, the Fe-$3d_{z^2}$ and Fe-$3d_{xy}$ orbitals characters become stronger around the $Z$ point, which is slightly different from other iron-based materials\cite{Boeri}. The inner cylindrical hole pocket is mainly contributed by Fe-$3d_{x^2-y^2}$ around zone center. It is interesting to point that the three-dimensional pocket around $Z$ point is mainly coming from the As-$4p_z$ orbital character (partial Fe-$3d_{z^2}$ orbital character), which is similar to the one in FeTe\cite{ASubedi} and KFe$_2$Se$_2$\cite{ZYLu}. The orbitals characters of band structures exhibit the typical characteristics of layered structures.

We now turn to discuss the superconducting pairing symmetry in KFe$_2$As$_2$.
Because of the the strong hole doping in KFe$_2$As$_2$, the electron pockets at $M$ point are replaced small hole pockets, and the hole pockets at $\Gamma$ point are very large in the folded BZ of two iron atoms per unit cell, as shown in Fig. \ref{fig:fig0}(b). The nesting between $\Gamma$ point and $M$ point is absent. Therefore, we start with a $t$-$J_1$-$J_2$-type model:
\begin{eqnarray}
\hat{H} &=& \sum_{i,\mu,\sigma} \varepsilon_{\mu}n_{i\mu\sigma} + \sum_{i,j}\sum_{\mu,\nu,\sigma} t_{ij}^{\mu\nu}\hat{c}^{\dag}_{i\mu\sigma}\hat{c}_{j\nu\sigma} \nonumber \\
 &&+ J_{1}\sum_{\langle i,j\rangle,\mu}(\vec{S}_{i\mu} \cdot \vec{S}_{j\mu} - \frac{1}{4}n_{i\mu}n_{j\mu}) \nonumber \\
 &&+ J_{2}\sum_{\langle\langle i,j\rangle\rangle,\mu}(\vec{S}_{i\mu} \cdot \vec{S}_{j\mu}- \frac{1}{4}n_{i\mu}n_{j\mu})
\label{EQ:Hamil}
\end{eqnarray}
where $i,j$ denote the sites in the square lattice of FeAs layers [see Fig. \ref{fig:fig0}(b)] and $\mu$, $\nu$ the orbitals, and $t_{ij}^{\mu\nu}$ is the transfer energy obtained from the maximally localized Wannier orbitals\cite{wannier1,wannier2,TAMaier}. As shown in Fig. \ref{fig:fig3}, the ten-orbital tight-binding band structure reproduces the density functional theory band structure rather accurately.
The last two terms represent exchange interactions between Fe-$3d$ electron spin with parameters
$J_1$ and $J_2$ denoting intralayer nearest-neighboring (n.n.) and next-nearest-neighboring (n.n.n.) sites exchange interaction, respectively. The symbols $\langle i,j\rangle$ and $\langle\langle i,j\rangle\rangle$ denote the summation over the n.n. and n.n.n. sites, respectively. The $\vec{S}_{i\mu}=\frac{1}{2}\sum_{\sigma,\sigma'}\hat{c}^{\dag}_{i\mu\sigma}\vec{\sigma}_{\sigma,\sigma'}\hat{c}_{i\mu\sigma'}$ and $n_{i\mu}=\sum_{\sigma}\hat{c}^{\dag}_{i\mu\sigma}\hat{c}_{i\mu\sigma}$ with $\vec{\sigma}$ representing the Pauli matrices operating on the spin indices. For the purpose to discuss the superconducting pairing symmetry, we obtain the following BCS-type Hamiltonian within the mean-field approximation:

\begin{figure}[tbp]
\includegraphics[bb=70 50 730 540, width=8.8cm, height=7.0cm]{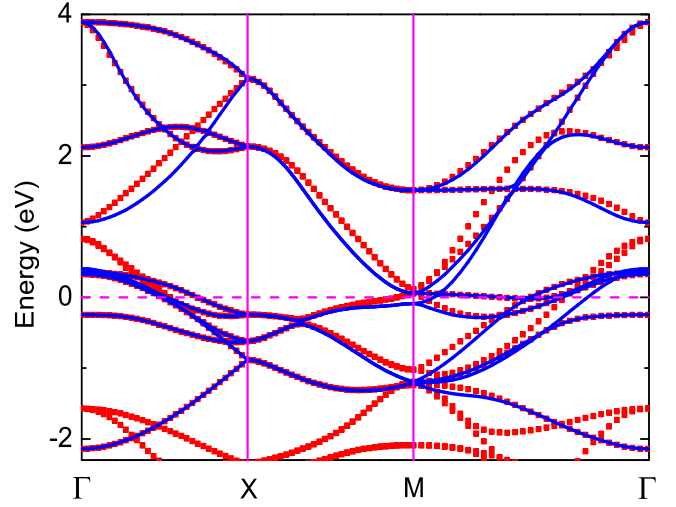}
\caption{(Color online) The paramagnetic density-functional theory band structure (red dot lines) and a Wannier fit (blue lines) of the ten orbitals in the vicinity of the Fermi surface onto the Fe-$3d$ orbitals. The Fermi energies are all set to zero.}\label{fig:fig3}
\end{figure}

\begin{eqnarray}
\hat{H} = \sum_{k,\mu,\nu,\sigma} \varepsilon_{k\mu\nu}\hat{c}^{\dag}_{k\mu\sigma}\hat{c}_{k\nu\sigma}+\sum_{k,\mu}\Delta_{\mu}(k)(\hat{c}^{\dag}_{k\mu\uparrow}\hat{c}^{\dag}_{-k\mu\downarrow}+h.c.)
\label{eq:two}
\end{eqnarray}

The superconducting order parameters satisfy the following BCS equations in the folded BZ with two iron atoms per unit cell,
\begin{subequations}
\begin{equation}
\Delta_{1,\mu}(k) =-\frac{J_1}{2} \sum_{k'}\phi^{1}_{k}\phi^{1}_{k'}<\hat{c}_{-k'\mu\downarrow}\hat{c}_{k'\mu\uparrow}>,\\
\end{equation}
\begin{equation}
\Delta_{2,\mu}(k) =-\frac{J_2}{2} \sum_{k'}\phi^{2}_{k}\phi^{2}_{k'}<\hat{c}_{-k'\mu\downarrow}\hat{c}_{k'\mu\uparrow}>.
\end{equation}
\label{eq:three}
\end{subequations}
Here the n.n. and n.n.n. intraorbital pairing symmetry factors should have the forms $\phi^{1}_{k}=4\cos \frac{k_x}{2} \cos \frac{k_y}{2}$ ($s_{x^2y^2}$-wave)
or $4\sin \frac{k_x}{2} \sin \frac{k_y}{2}$ ($d_{xy}$-wave), and $\phi^{2}_{k}=2(\cos k_x + \cos k_y)$ ($s_{x^2+y^2}$-wave) or $2(\cos k_x - \cos k_y)$ ($d_{x^2-y^2}$-wave) in folded BZ with two iron atoms per unit cell.
Note that the $d_{xy}$-wave and $d_{x^2-y^2}$-wave pairing symmetries are nodal and $s_{x^2y^2}$-wave and $s_{x^2+y^2}$-wave are nodeless for any small doping parameter. In a previous study of a multiorbital $t$-$J_1$-$J_2$ model for iron pnictides\cite{KSeo,Goswami,RYu,YHSu}, it has been demonstrated that the dominant pairing symmetry is governed by the intraorbital exchange interactions, and the interorbital exchange interaction only introduces quantitative modifications of the phase
diagram. Therefore to keep the analysis simple, we will only consider the intraorbital pairings. The pairing order parameter $\Delta_{\mu}(k)$ for the orbital $\mu$ is linear combinations of four intraorbital pairings $\Delta_{i,\mu}(k)$.

\begin{figure}[tbp]
\includegraphics[bb=1 1 730 530, width=8.5cm, height=7cm]{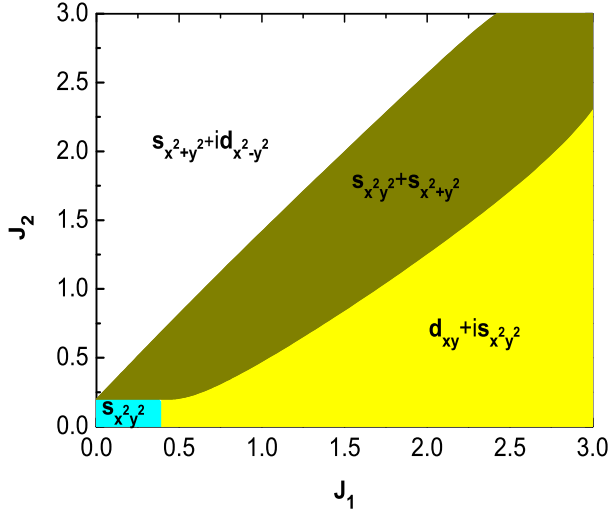}
\caption{(Color online) Zero temperature superconducting phase diagram in the $t$-$J_{1}$-$J_2$ plane obtained by solving the effective model in Eqs. (\ref{eq:two}).}\label{fig:fig4}
\end{figure}

The above equations: Eq.(\ref{eq:two}) and Eq.(\ref{eq:three}) can be solved self-consistently, we obtain the quasiparticle dispersion spectra $E_{k\mu}$. The pairing gap matrix is determined by minimizing the ground state energy density\cite{Goswami,RYu}: $f=\sum_{i,k,\mu}\frac{J_i}{4}|\phi^{i}_{k}<\hat{c}_{-k\mu\downarrow}\hat{c}_{k\mu\uparrow}>|^2-\sum_{k,\mu}(E_{k\mu} - \varepsilon_{k}^{\mu})$, where $\varepsilon_{k}^{\mu}$ is the eigenvalue of ten-orbital tight-binding model.

When the effective kinetic energy term is absent, the $J_1$ dominating $s_{x^2y^2}$-wave and $d_{xy}$-wave pairing symmetries are indeed degenerate. Likewise, the $s_{x^2+y^2}$-wave and $d_{x^2-y^2}$-wave dominated by $J_2$ are energetically degenerate, and all four pairing states mix together when $J_2$ is comparable with $J_1$. Taking the effective kinetic energy term into account, the degeneracy states will be lefted, and the phase diagram of pairing symmetries is shown in Fig. \ref{fig:fig4}. The phase on the left upper corner has a time reversal breaking phase with $s_{x^2+y^2} + id_{x^2-y^2}$-wave dominated by n.n.n. intraorbital pairing interaction $J_2$. When the n.n. pairing interaction $J_1$ increases and is comparable to the n.n.n. pairing interaction $J_2$, the $d_{x^2-y^2}$-wave will be substituted by $s_{x^2y^2}$-wave and lead to a $s_{x^2+y^2}+s_{x^2y^2}$-wave in the middle part of Fig. \ref{fig:fig4}. When the n.n. pairing interaction $J_1$ further increases, the $s_{x^2y^2}$-wave will be suppressed and substituted by $d_{xy}$-wave and lead to another kind time-reversal symmetry breaking state $s_{x^2y^2}$+$id_{xy}$-wave. On the left lower corner of the phase diagram, when n.n. pairing interaction $J_1$ is dominating and has a very small value, the system favors a nodeless $s_{x^2y^2}$-wave state.

\begin{figure}[tbp]
\includegraphics[bb=50 5 730 620, width=8cm, height=7cm]{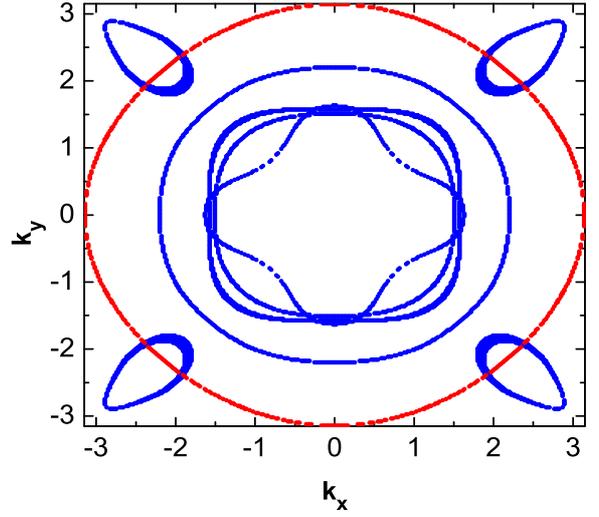}
\caption{(Color online) A schematic plot for the gap nodal structure in the folded BZ with two irons per unit cell. The blue lines denote the FS topology of the ten-orbital model. The red line is the contour line for zero gap value of $s_{x^2+y^2}+s_{x^2y^2}$-wave order parameter in Eq.(\ref{eq:four}) with $\delta = 0.2$.}\label{fig:fig5}
\end{figure}

From the view point of experiments, the nodes have been observed by serval groups\cite{Furukawa,JKDong,Terashima,Reid,Hashimoto,Fukazawa,SWZhang,Ohishi}. Therefore, we conclude that only the $s_{x^2+y^2}+s_{x^2y^2}$-wave can give rise to the nodal structure, and only this phase can explain the experiments observed nodes based on the aforementioned discussions. The order parameter in this mixed phase can be written as

\begin{equation}
\Delta_{\mu}(k) = \Delta_0[\cos \frac{k_x}{2} \cos \frac{k_y}{2} + \delta (\cos k_x + \cos k_y)]\label{eq:four}
\end{equation}
where the value $\Delta_0$ is a constant and $\delta$ depends on the n.n. interorbital pairing interaction $J_1$ and n.n.n. intraorbital pairing interaction $J_2$. When the n.n. intraorbital pairing interaction $J_1$ is comparable to the n.n.n. one $J_2$, the gap zero points will develop. As shown in Fig. \ref{fig:fig5}, when $\delta$ reaches to a certain value, the contour of the gap zeros will cross the FS topology, which leads to the nodal behavior. This result can be tested by future experiments, such as APRES measurements.

In conclusion, we have constructed a ten-orbital model by using the maximally localized Wannier functions based on the first-principles band structure calculations and combined with the $J_1$-$J_2$ model, all possible pairing symmetries are examined by solving the self-consistently BCS equations. The mean-field phase diagram of a $t$-$J_1$-$J_2$ model for this compound is constructed. We find that only the $s_{x^2+y^2}+s_{x^2y^2}$-wave pairing symmetry can give rise to the nodal structure, and only this wave pairing symmetry can explain that experiments which evidenced a nodal order parameter. This result can be tested by future experiments.

\begin{acknowledgments}
This work was supported by the Texas Center for Superconductivity at the University of Houston and by the Robert A. Welch Foundation under Grant No. E-1146 (W.L., J.L. and C.S.T.), and by the
National Nuclear Security Administration of the U.S. Department of
Energy at LANL under Contract No. DE-AC52-06NA25396 and the U.S. DOE
Office of Basic Energy Sciences (J.-X.Z.). Y.C. was supported by the National Natural Science Foundation of China (Grant No. 11074043) and the State Key Programs of China (Grant Nos. 2009CB929204 and 2012CB921604). W.L. also gratefully acknowledges the financial support by Research Fund of Fudan University for the Excellent Ph.D. Candidates, and thanks R.B. Tao, J.P. Hu, S.H. Pan, Z. Fang, H.J. Xiang, D.G. Zhang and G. Xu for great helpful discussions.
\end{acknowledgments}

\end{document}